# Activated Vibrational Modes and Fermi Resonance in Tip-Enhanced Raman Spectroscopy

Mengtao Sun, [1] Yurui Fang,[1] Zhenyu Zhang [2,3] and Hongxing Xu [1,4,*]

[1] *Beijing National Laboratory for Condensed Matter Physics, Institute of Physics, Chinese Academy of Sciences, P. O. Box 603-146, Beijing, 100190, People's Republic of China*

[2] *ICQD, Hefei National Laboratory at the Microscale, University of Science & Technology of China, Hefei, People's Republic of China*

[3] *Department of Physics and Astronomy, University of Tennessee, Knoxville, TN 37996-1200*

[4] *Division of Solid State Physics, Lund University, Lund 22100, Sweden*

**ABSTRACT**

Using p-aminothiophenol (PATP) molecules on a gold substrate as prototypical examples and high vacuum tip-enhanced Raman spectroscopy (HV-TERS), we show that the vibrational spectra of those molecules are distinctly different from those in typical surface-enhanced Raman spectroscopy. Detailed first-principles calculations help to assign the Raman peaks in the TERS measurements as Raman active and infrared (IR) active vibrational modes of dimercaptoazobenzene (DMAB), thus providing strong spectroscopic evidence for the conversion of PATP dimerization to DMAB. The activation of the IR active modes is due to enhanced electromagnetic field gradient effects within the gap region of the highly asymmetric tip-surface geometry. Our TERS measurements also realize splitting of certain vibrational modes due to Fermi resonance between a fundamental mode and the overtone of a different mode or a combinational mode. These findings help to broaden the versatility of TERS as a promising technique for ultrasensitive molecular spectroscopy.

First demonstrated by Stöckle et al. over a decade ago [1], tip-enhanced Raman spectroscopy (TERS) is a high-sensitivity optical analytical technique with high spatial resolution beyond the diffraction limit of light. In TERS, a sharp metal tip is used to create a "hot site" to excite localized surface plasmons and consequently enhance the electromagnetic field and Raman signals in the vicinity of the tip apex [1-6]. The tip can be moved three dimensionally to control the position of the "hot site" and the corresponding enhancement factor by changing the gap distance between the tip and the substrate. Therefore, TERS has the inherent advantage of overcoming one of the most severe restrictions in the application of surface enhanced Raman scattering (SERS), which usually requires roughness of metal surfaces or aggregations of metal nanoparticles to create "hot sites" that are hardly controllable. TERS may solve a wide variety of problems in high vacuum (HV) single crystal surface science, electrochemistry, heterogeneous catalysis, microelectronics, and tribology, thus offering new opportunities for gaining insights in the physics and chemistry of these diverse systems.

As a compelling and challenging example, the appearance of normally unseen Raman modes in the spectrum of p-aminothiophenol (PATP) adsorbed on metal surfaces has been a long-standing issue in SERS. Earlier explanations often ascribe this phenomenon to selective enhancement of some $b_2$ modes of PATP due to charge transfer between the molecules and the metal substrates [7-10]. Recently, the dimerization of PATP to dimercaptoazobenzene (DMAB) has been suggested to interpret these unseen Raman modes [11-15]. Both theoretical simulations [11-12] and experimental observations [12-13] offer credible support of the explanation based on the dimerization mechanism. However, due to the broadening and relatively few Raman modes in the SERS spectra, controversies remain about the dimerization explanation [16, 17]. Separately, it has been suggested that the gradient effect associated with the strongly enhanced electric field at a rough metal surface [18-20] could activate IR-active modes, which are usually Raman inactive. If the IR-active modes of DMAB are also observed, together with the Raman-active modes in SERS studies, it is expected that the dimerization picture will gain more convincing spectroscopic evidences.

In this Letter, by exploring the advantages of TERS, we investigate this long-standing issue, the appearance of new Raman modes of adsorbed PATP molecules that are typically unseen in normal Raman spectroscopy. Much narrower and many more Raman peaks are observed in our high vacuum TERS (HV-TERS) system than in previous SERS studies. By comparing with the results of our density function theory simulations, all these Raman peaks can be assigned as Raman-active symmetric modes and IR-active asymmetric modes of the DMAB molecule. The fully activated and strongly enhanced IR-active modes observed in HV-TERS here give more convincing spectroscopic evidences to support the conversion of PATP via dimerization into the putative product of DMAB. The gradient effect associated with the enormously enhanced electric field within the nanogap region between the sharp gold tip and the gold film can activate the IR-active modes of the DMAB molecule to be Raman-active. The huge electromagnetic enhancement in the nanogap region further results in distinct Fermi resonances, characterized by strong coupling of a fundamental mode and a overtone of a different mode or a combinational mode, to split the corresponding Raman modes.

TERS spectra were measured with a home-built HV-TERS setup. The schematic diagram of the HV-TERS system is shown in Fig. 1. It consists of a homemade scanning tunneling microscope (STM) in a high vacuum chamber, a Raman spectrometer combined with a side illumination of 632.8 nm He-Ne laser light with an angle of $60^o$ for Raman measurements, and three-dimensional piezo stages for the tip and sample manipulations. The pressure in the chamber is ~$10^{-7}$ Pa. A gold tip with a ~50 nm radius was made by electrochemical etching of a 0.25 mm diameter gold wire [21]. The substrate was prepared by evaporating a 100 nm gold film to a newly cleared mica film under high vacuum. The film was immersed in a $1\times10^{-5}$ M PATP ethanol solution for 24 hours, then washed with ethanol for 10 minutes to guarantee that there was only one monolayer of PATP molecules adsorbed on the gold film. Then the sample was put into the high vacuum chamber immediately. To get a good signal-to-noise ratio, the TERS signals were collected with an acquisition time of 60 seconds and accumulated 20 times for each spectrum.

We measured TERS spectra of PATP adsorbed on the Au film at different bias voltages and currents, and found the optimal conditions to be ± 1 V (bias voltage is on the sample) and 1 nA (current) in our HV-TERS system. Two series of typical measured TERS spectra at the above experimental conditions for different biases are shown in Figs. 2(a) and 2(b), respectively. It is found that at these conditions, the fluctuations of these TERS spectra are small, and the TERS spectra are stable and can be readily repeated experimentally. The profiles of the TERS spectra obtained at +1 V and -1 V are quite similar, indicating that there is not much influence from the polarity of the bias voltage on the TERS spectra. For comparison, the typical SERS spectrum of PATP in Au sol and the normal Raman scattering (NRS) spectrum of PATP powder (Fig. 2(c)) were also measured with a Renishaw inVia Raman system and excited with light at 632.8 nm. The molecular structure of DMAB and PATP are also shown in the insets of Fig. 2(c), respectively.

Comparing Figs. 2(a) and 2(b) with Fig. 2(c), it is found that the TERS spectra of PATP adsorbed on the Au film are significantly different from the SERS and NRS spectra of PATP. Much narrower and many more Raman peaks are observed in Figs. 2(a) and 2(b) for the HV-TERS study than in Fig. 2(c) for the SERS and NRS studies. It is worthwhile to note that our previous report showed that the SERS spectra of PATP adsorbed on a bare Au film is very similar to the NRS spectrum of PATP [15], whereas SERS of PATP on Au colloidal nanoparticles is different from the corresponding NRS spectrum. The later case has been attributed to the plasmon-assisted chemical reaction of PATP dimerization into DMAB, but not the case for the bare Au film where the plasmon enhancement is weak. The rich spectroscopic features in TERS could also suggest the chemical reaction of PATP dimerization due to strong local electromagnetic enhancement within the nanogap region of the tip-surface geometry. Moreover, the polarity of the bias voltage does not influence the profiles of the TERS spectra in Fig. 2(a) and Fig. 2(b), which may indicate that the detected molecule should be structurally symmetric, thus offering additional evidence for the formation of DMAB converted from PATP to symmetrically bridge two metal surfaces (the Au tip and the Au film) with two thiol groups. But the difference between TERS and SERS spectra remains.

To correctly interpret the above spectral phenomena, we choose two TERS spectra at different biases for vibrational mode assignments as shown in Figs. 3(a) and 3(b). Furthermore, the Raman spectra of DMAB were calculated within density functional theory [22] using the Gaussian 09 suite [23] with PW91PW91 functional [24], a 6-31G(d) basis set for C, N, S and H, and a LANL2DZ basis set [25] for Au, where a $Au_5$-DMAB-$Au_5$ junction was employed to simulate the Raman spectra mimicking the TERS spectra. The IR spectra of DMAB in the $Au_5$-DMAB-$Au_5$ junction were also simulated with the same method. Based on the simulated Raman spectrum in Fig. 3(c), all the Raman-active symmetric ag vibrational modes can be assigned in Figs. 3(a) and 3(b). Interestingly, most of the remaining Raman peaks in Figs. 3(a) and 3(b) can be assigned as IR-active asymmetric bu modes according to the simulated IR spectrum in Fig. 3(d). It has been previously studied that the gradient-field effect can activate IR-active modes in SERS due to molecular quadrupole transitions when the molecules are placed near metal surfaces [18-19]. Such gradient-field effects were also observed in near-field optical microscopy Raman (NSOM-Raman) [20]. Here, with similar reasons, the observation of the full activation of IR-active modes should be caused by very high field gradients within the nanogap of a highly asymmetric geometry consisting of a sharp metal tip and a flat metal surface in our HV-TERS system. Moreover, the enormous electromagnetic field enhancement in the nanogap region makes this phenomenon of activation IR-active modes more prominent to be easily observed.

Next we compare the simulated Raman spectrum in Fig. 3(c) and the SERS spectrum in Fig. 2(c). It is found that six strong peaks in the experimental and simulated results without scaling are consistent well with each other. We also note that in the simulated Raman spectrum of DMAB in Fig. 3(c), the wavenumbers have been scaled by a factor of 1.014 (or blue shifted by ~1%) for closer comparison with the TERS spectra shown in Figs. 3(a) and 3(b). Such blue shifts could be easily caused by weak external perturbations, such as tip, current, and voltage in the HV-TERS system.

Figure 4(a) shows a zoom-in spectrum of Fig. 3(a), where each of the symmetric and asymmetric C-N stretching modes assigned as the Raman-active $ag_{13}$ mode and IR-active asymmetric $bu_{13}$ mode, respectively, splits into two Raman peaks. The splitting of Raman peaks results from Fermi resonance (FR). As a result of a FR, an overtone of a different fundamental mode or a combination mode can appear in the vibrational spectra by gaining spectral weight from a fundamental mode [26]. Fermi resonances are frequently found in IR or Raman spectra in symmetric triatomic molecules such as $CO_2$ and $CS_2$ [26-28], but it is the first observation of FR in the TERS study reported here. When the molecules are placed in the nanogap region, external perturbations from enormously enhanced electromagnetic fields, high field gradients, and the Au tip could cause such distinct Fermi resonances.

For a Fermi resonance, the split vibrational energies under the perturbation can be written as [26-28],

$$E_\pm = \frac{1}{2}(E_A + E_B) \pm \frac{1}{2}\sqrt{(E_A - E_B)^2 + 4\phi^2} , \qquad (1)$$

where $E_A$ and $E_B$ are the vibrational energies of the fundamental mode and an overtone of a different fundamental mode or a combination mode before splitting, respectively, and $\phi$ is the FR coupling coefficient, which describes the coupling strength of the fundamental vibrational mode and the combinational mode (or the overtone mode) in the Fermi resonance. When the FR coupling coefficient is larger, the spectral energy splitting between the fundamental vibrational mode and the combinational mode (or the overtone mode) is larger as well. In addition, when the unperturbed frequencies $E_A$ and $E_B$ in Eq. (1) is closer, the FR coupling coefficient becomes larger. For our case in Fig. 4(a), the unperturbed frequencies $E_A$ and $E_B$ for the Raman-active symmetric ag$_{13}$ or the IR-active asymmetric bu$_{13}$ can be estimated as:

$$E_A = \frac{E_+ + E_-}{2} + \frac{E_+ - E_-}{2} \times \frac{I_+ - I_-}{I_+ + I_-} \qquad (2)$$

$$E_B = \frac{E_+ + E_-}{2} - \frac{E_+ - E_-}{2} \times \frac{I_+ - I_-}{I_+ + I_-} , \qquad (3)$$

where $I_+$ and $I_-$ are the Raman intensities of the split peaks, respectively. The difference in energy between the perturbed levels in the presence and absence of Fermi resonance can be obtained with $\Delta E_\pm = E_+ - E_-$ and $\Delta E_{AB} = |E_A - E_B|$, respectively.

In Fig. 4(a), we obtain $I_+(ag_{13}) \approx I_-(ag_{13})$ and $I_+(bu_{13}) \approx \frac{1}{2}I_-(bu_{13})$, $\Delta E_\pm(ag_{13}) = 20.4$ cm$^{-1}$ and $\Delta E_\pm(bu_{13}) = 14.6$ cm$^{-1}$. For the Raman-active symmetric ag$_{13}$ mode, according to Eqs. 2 and 3, we obtain $E_A = E_B = \frac{E_+ + E_-}{2}$, i.e. $\Delta E_{AB} \approx 0$, and the FR coupling coefficient can be calculated by Eq. 1 as $\phi(ag_{13}) = \frac{1}{2}\Delta E_\pm(ag_{13}) = 10.2$ cm$^{-1}$. As illustrated in Figs. 4(b) and 4(c) with different vibrational modes marked as vertical short lines, it can be seen that the fundamental Raman-active symmetric ag$_{13}$ mode at 1213 cm$^{-1}$ is very close to the combinational mode of ag$_6$ at 727 cm$^{-1}$ and bg$_6$ at 485 cm$^{-1}$ with $\Delta E_{AB} = E(ag_{13}) - [E(bg_6) + E(ag_6)] \approx 1$ cm$^{-1}$. The symmetry of the combinational mode $bg \times ag = bg$ is asymmetric.

Similarly, for the IR-active asymmetric $bu_{13}$ mode, we obtain $E_A = \frac{E_+ + E_-}{2} + \frac{1}{3} \times \frac{\Delta E_\pm}{2}$ and $E_B = \frac{E_+ + E_-}{2} - \frac{1}{3} \times \frac{\Delta E_\pm}{2}$, according to Eqs. 2 and 3, and $I_+(bu_{13}) \approx \frac{1}{2}I_-(bu_{13})$ as in Fig. 4(a). Then, we can obtain $\Delta E_{AB}(bu_{13}) = \frac{1}{3}\Delta E_\pm(bu_{13}) = 4.9$ cm$^{-1}$ and $\phi(bu_{13}) = \frac{\sqrt{2}}{3}\Delta E_\pm(bu_{13}) = 6.88$ cm$^{-1}$. As illustrated in Figs. 4(b) and 4(c), the calculated IR-active asymmetric bu$_{13}$ mode at 1267 cm$^{-1}$ is about 7 cm$^{-1}$ smaller than the combinational mode of $bu_5$ at 634 cm$^{-1}$ and $bu_6$ at 640 cm$^{-1}$. The difference between experimental and theoretical results (4.9 cm$^{-1}$ and 7 cm$^{-1}$) is about 2 cm$^{-1}$. The symmetry of the combinational mode is $bu \times bu = ag$, which is symmetric.

In summary, by using the HV-TERS system, we have demonstrated the ready conversion of PATP dimerization to DMAB, possibly catalyzed by the presence of the TERS tip on top of the gold film. The definitive observation of the IR-active bands associated with the DMAB molecules in the TERS spectra provides further experimental evidence for the dimerization mechanism. The activation of these extra modes is primarily attributed to enhanced electromagnetic field gradient effects within the gap region of the tip-surface geometry. Moreover, strongly enhanced plasmon responses in the nanogap region further results in distinct Fermi resonance between a fundamental vibrational mode and the overtone of a different vibrational mode or a combinational mode to split the corresponding Raman peaks. Aside from the fundamental significance, these original findings also help to significantly broaden the capacity of TERS as an analytical tool for ultrasensitive molecular spectroscopy.

**ACKNOWLEDGMENT:** This work was supported by the National Natural Science Foundation of China (Nos. 10874233, 10874234, 90923003, 20703064 and 11034006), the National Basic Research Project of China (Grants 2009CB930701, 2009CB930704 and 2007CB936804), the U.S. National Science Foundation (Grant No. 0906025), the U.S. Department of Energy, Basic Energy Sciences, Materials Sciences and Engineering Division, and the Nanometer Structure Consortium at Lund University, nmC@LU (H.X.X).

**Figure captions**

**Figure 1.** (color on line) Schematic diagram of the home-built HV-TERS setup.

**Figure 2.** (a) and (b) TERS spectra of DMAB at experimental conditions of 1 nA current, +1 V, and -1 V bias voltage, respectively, measured at different positions. (c) SERS spectrum of DMAB in Au sol and the normal Raman scattering spectrum of PATP.

**Figure 3.** (color on line) (a) and (b) TERS spectra of DMAB at experimental conditions of 1 nA current, -1 V and +1 V bias voltage, respectively. (c) Simulated Raman and (d) IR spectra of DMAB, where the wavenumber is scaled by 1.014.

**Figure 4.** (color on line) (a) The experimentally observed Fermi resonance and simulated TERS spectra of DMAB for Raman-active symmetric $ag_{13}$ and IR-active asymmetric $bu_{13}$ modes. (b) and (c) The calculated vibrational modes and the combinational modes for Fermi resonance, where the wavenumber is scaled by 1.014.

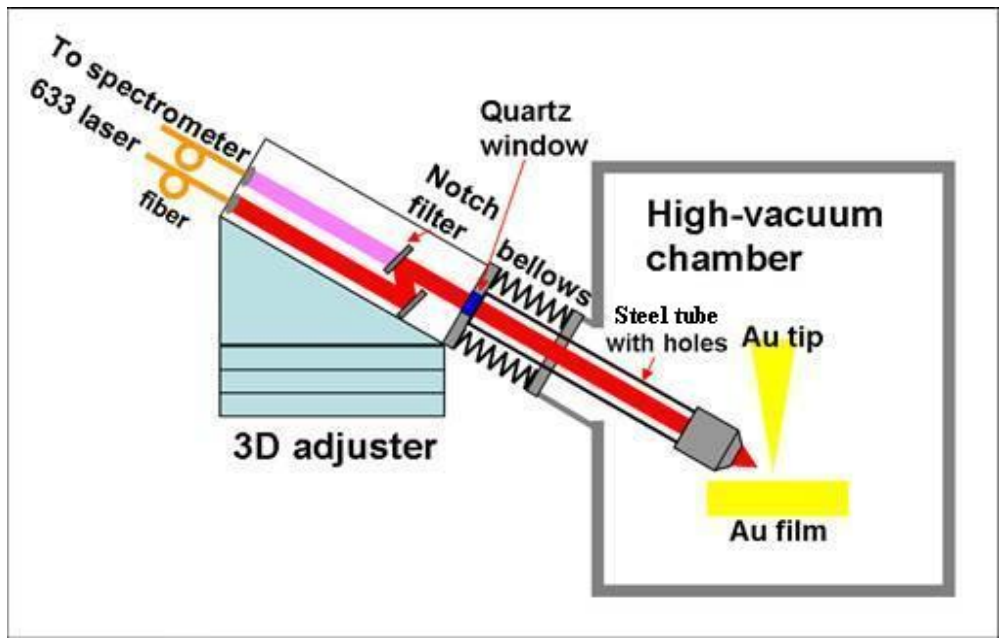

Figure 1. Sun et al.

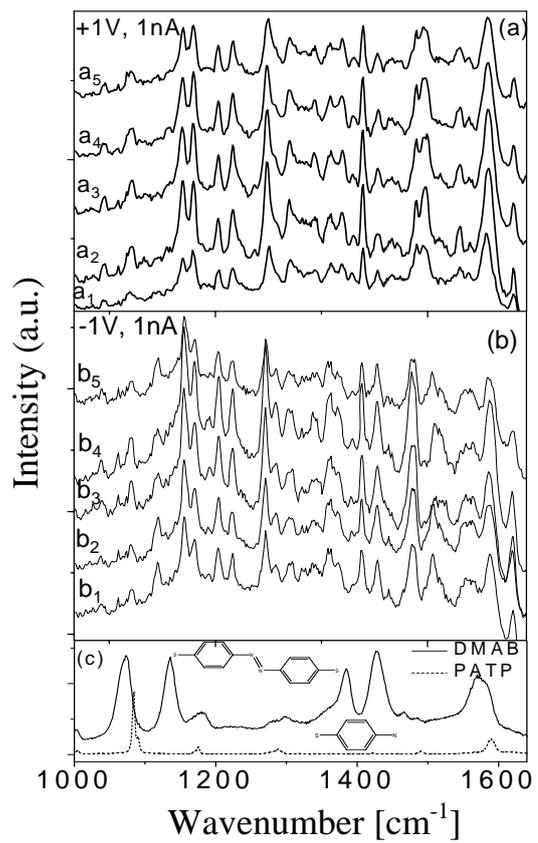

Figure 2. Sun et al.

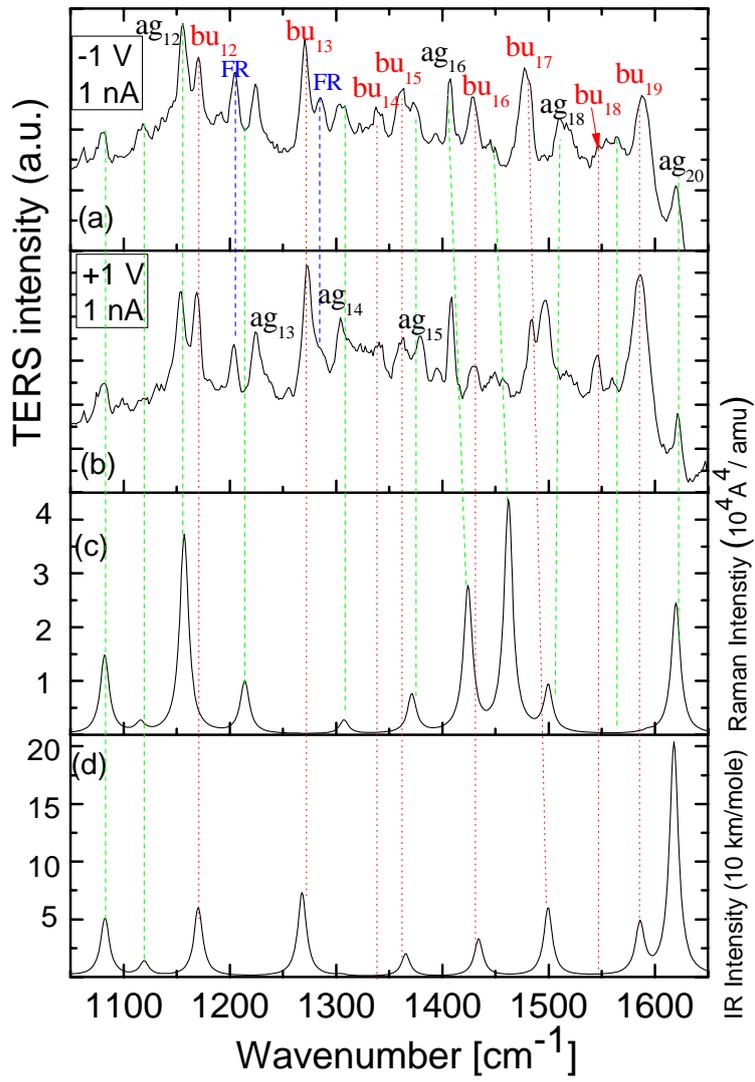

Figure 3. Sun et al.

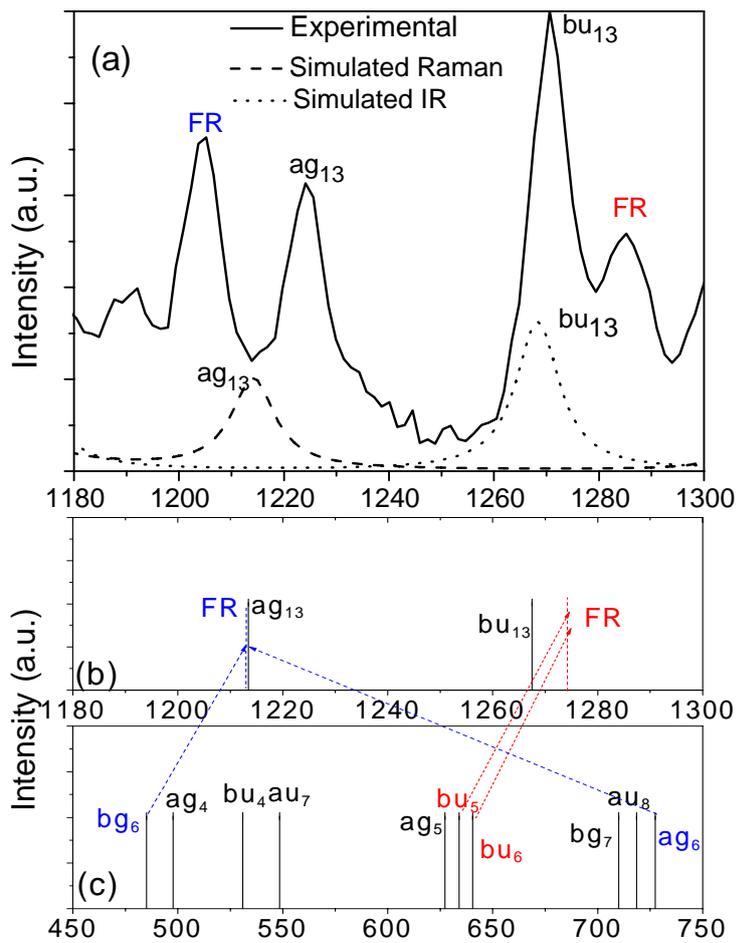

Figure 4. Sun et al.